\documentclass[a4paper,11pt]{article}
\usepackage{jheppub}
\usepackage[T1]{fontenc}
\usepackage{graphicx}
\usepackage{amsmath}
\usepackage{amsmath}
\usepackage{xspace}
\usepackage{color}
\usepackage[utf8]{inputenc}
\usepackage{commath}
\usepackage{slashed}
\usepackage{fancyvrb}
\usepackage{braket}
\usepackage{multirow}
\usepackage{color,soul}
\usepackage{appendix}
\usepackage{cleveref}

\title{Strongly Interacting Dark Matter admixed Neutron Stars}

\author{Yannick Dengler,}
\author{Suchita Kulkarni,}
\author{Axel Maas,}
\author{Kevin Radl}
\affiliation{Institute of Physics, NAWI Graz, University of Graz, Universit\"atsplatz 5, 8010 Graz, Austria}
\emailAdd{yannick.dengler@uni-graz.at}
\emailAdd{suchita.kulkarni@uni-graz.at}
\emailAdd{axel.maas@uni-graz.at}
\emailAdd{kevin.radl@edu.uni-graz.at}

\abstract{Dark matter may accumulate in neutron stars given its gravitational interaction and abundance. We investigate the influence of strongly-interacting dark matter, described by a QCD-like one-flavor $G_2$ gauge theory, on neutron stars. This choice allows to test, for the first time, a first-principles-determined non-Abelian dark matter equation of state, which supports composite fermionic dark matter and thus a Fermi-pressure-stabilized dark matter component. The ordinary matter part of the mixed star is described by available model-agnostic equations of state that interpolate between the low-density regime and high-density regime. We find that strongly-interacting dark matter has a similar impact on neutron stars as other model equation of states and confirm that strongly-interacting dark matter can be accommodated by constraints from neutron star observations within our uncertainties.}
\begin{document}
\maketitle
\flushbottom


\section{Introduction}

The dark matter hypothesis~\cite{Rubin:1978kmz,Bambi:2015mba,Cirelli:2024ssz} is and remains one of the primary candidates to resolve a multitude of astronomical observations, despite a lack of direct observation. Assuming the existence of dark matter, it interacts gravitationally by definition. In addition, indirect evidence is compatible with a sizable dark matter self-interaction~\cite{Spergel:1999mh,Kaplinghat:2015aga,Tulin:2017ara,Sagunski:2020spe,Andrade:2020lqq,Eckert:2022qia,Bose:2022obr,Adhikari:2022sbh, Salucci:2018hqu}, while interactions with standard model matter (in the following called \textit{ordinary}) remain only bounded from above~\cite{Bambi:2015mba,Cirelli:2024ssz}. Concentrating on dark matter gravitational signals is therefore a prudent strategy. However, except for supermassive dark matter particles~\cite{Carr:2020gox}, individual dark matter particles will not create an observable gravitational signal. Thus, only accumulations of dark matter particles will provide relevant signals, requiring an accumulator.

Particle dark matter will necessarily accumulate in neutron stars making them an excellent laboratory to study dark matter gravitationally on small scales~\cite{Akmal:1998cf,Lattimer:2000nx,Ozel:2016oaf}. The possible amount of dark matter may vary depending on the age, history, and location of the neutron star, but most estimates conclude that that it is very unlikely to exceed 1\% of the total neutron star mass~\cite{Guver:2012ba,Giangrandi:2022wht,Giangrandi:2025rko}, though numbers as large as up to 10\% have also been motivated \cite{Grippa:2024ach,DelPopolo:2020hel}. Nevertheless, a careful analysis of potential effects of dark matter on gravitational and electromagnetic signals from neutron stars is necessary to assess if, and if yes what, impact dark matter could have on its properties. This will allow to understand whether neutron stars can be used to learn about dark matter. It also will clarify whether dark matter could, potentially at least, create effects relevant to understanding the ordinary matter aspects of neutron stars. Both aspects require a good data basis on neutron stars.

On the observational side, fortunately, we do have a very good overview of global neutron star properties from observations. Among the direct observations, the Neutron star Interior Composition ExploreR (NICER) mission is dedicated to direct observation of neutron star properties via observations in the soft (0.2–12 keV) X-ray band~\cite{Gendreau2024}. The mission concentrates on measurements of neutron star mass and radius, and will in its runtime provide a wide range of characterized neutron stars. Thus, a stochastically large enough sample is available to also detect variations of neutron star properties due to varying amounts of dark matter admixtures. In contrast to a single-component neutron star of ordinary matter, which yields a unique\footnote{There can, of course, be variation due to temperature (and thus age), chemical composition in the crust and atmosphere, and further effects, also fuzzing out the mass-radius relation of pure ordinary matter neutron stars. These effects can, at least in principle, be quantified and accounted for. Any additional dark matter components will add on these effects, and will yield a deviation.} mass-radius relation, differing amounts of dark matter will fuzzy out such a curve. Thus, first-principle knowledge of this fuzzying will allow to detect dark matter accumulation.

Moreover, neutron stars with a large amount of dark matter in the interior, also known as mixed neutron stars, might provide smoking gun signals like supplementary peaks in the gravitational wave power spectral density~\cite{Giangrandi:2022wht,Ellis:2017jgp}. Mixed neutron star binaries are thus a further testbed for dark matter properties, as they are capable of generating exotic signals for gravitational waves experiments. In particular, the so-called tidal deformability, indicating the deformation of colliding neutron stars, has been constrained by LIGO observations of a single binary neutron star merger GW170817~\cite{LIGOScientific:2017vwq,LIGOScientific:2017ync}. Along with this LIGO has also measured the ``chirp mass'' for the same system. These observations have been combined with simultaneous observations from other multi-messenger telescopes, leading to an even more comprehensive analysis~\cite{LIGOScientific:2017ync}.

On the theoretical side, a primary ingredient in understanding such observable effects is the equation of state, which determines the evolution of thermodynamic properties, for both the ordinary component and a potential dark matter component inside the neutron star. Owing to the sign problem~\cite{Gattringer:2010zz}, lattice QCD -- the methodological mainstay to determine much of QCD physics -- is unable to provide a fully controlled equation of state of ordinary matter inside neutron stars. Limited knowledge can be derived from heavy-ion experiments, models, and other methods~\cite{Friman:2011zz}. However, this has so far been insufficient to provide a final and complete understanding. In fact, observed neutron star properties are used to constrain the QCD equation of state~\cite{Akmal:1998cf,Lattimer:2000nx,Ozel:2016oaf,LIGOScientific:2017ync}, under the assumption that neutron stars are primarily composed of ordinary matter. Therefore, we will use three different candidates for the equation of state of ordinary matter to represent this uncertainty. The particular choices will be discussed in  \Cref{s:eos}. However, they are made to reduce model dependence as much as possible in favor of data-driven input.

Thus, while neutron stars may accumulate some amount of dark matter, and consequently could be dark matter discovery avenues, insufficient knowledge on the accumulated amount coupled with remaining uncertainties over neutron star ordinary physics present a challenging situation for the subject of dark matter admixed neutron stars. Progress with this conundrum would therefore be made if at least the dark matter equation of state would be controlled from first principles, reducing the uncertainties on the evolution of one of the two neutron star components. This is possible for weakly interacting massive particles (and similar) cases, and several studies have been performed in this regard~\cite{Quddus:2019ghy,Quddus:2020lbt}. There have also been numerous other studies, which have been based on various assumptions and simplifications of the details of the dark matter equation of state~\cite{Guver:2012ba,Biesdorf:2024dor,Hajkarim:2024ecp,Kunkel:2024otq,Wystub:2021qrn,Pitz:2024xvh,Diedrichs:2023trk,Dengler:2021qcq,Tolos:2015qra,Mukhopadhyay:2015xhs,Rezaei:2016zje,Panotopoulos:2017idn,Nelson:2018xtr,Ellis:2018bkr,Gresham:2018rqo,Ivanytskyi:2019wxd,Karkevandi:2021ygv,Sen:2021wev,Guha:2021njn,Goldman:2013qla,Xiang:2013xwa,Li:2012ii,Sandin:2008db,Leung:2011zz,Leung:2012vea,Barbat:2024yvi,Hippert:2021fch,Issifu:2024htq,Karkevandi:2024vov,Shakeri:2022dwg,Ema:2024wqr}.

Owing to the possible existence of large dark matter self-interactions~\cite{Spergel:1999mh,Kaplinghat:2015aga,Tulin:2017ara,Sagunski:2020spe,Andrade:2020lqq,Eckert:2022qia,Bose:2022obr,Adhikari:2022sbh, Salucci:2018hqu}, QCD-like hidden sectors have recently been in the spotlight as the resulting dark hadrons can inherently generate the required self-interactions. Such dark sectors thus demand their own equation of state calculations, should neutron stars be used to probe their existence. Similar to ordinary matter, these calculations may, however, also suffer from a sign problem. Fortunately, for some of the cases, like the Sp($N$)-based QCD-like theories ~\cite{Hochberg:2014dra,Hochberg:2014kqa,Kulkarni:2022bvh}, the sign problem is absent, and calculations may be possible, or have been done like for a dark matter model based on a two-color QCD-like hidden sector~\cite{Drach:2015epq}, see, e.\ g.,~\cite{Cotter:2012tt,Boz:2019enj,Hands:2024nkx}.

But, both, Sp($N$)-based and two-color based QCD-like strongly-interacting dark matter scenarios, feature only bosonic hadrons. Such dark matter systems do not experience stabilization by Fermi pressure, and thus gravitational collapse will only be halted when the dark quark substructure becomes relevant. This happens usually only at asymptotically high densities~\cite{Friman:2011zz}, yielding rather a dense dark matter nugget at the center of the neutron star at best, unless strong self-interactions provide stabilization~\cite{Liebling:2012fv}. 

Therefore any study of mixed neutron stars involving a new QCD-like admixture will benefit from a dark QCD-like sector with fermionic dark hadrons, where the equation of state is accessible in first-principle calculations. Fortunately, such possibilities exist, e.\ g.\,, a hidden QCD-like sector based on the gauge group G$_2$~\cite{Holland:2003jy,Maas:2012wr,Wellegehausen:2013cya}. It has already been shown that this theory can support pure compact stellar objects~\cite{Hajizadeh:2017jsw}, which could be interpreted as pure dark stars. Already the one-flavor theory provides a conserved quantum number carried by the fermionic dark baryons~\cite{Maas:2012wr}, and thereby offers a dark matter candidate, alongside a rich spectrum of further states~\cite{Wellegehausen:2013cya}. The two-flavor case furthermore offers dark pions as additional stable states with a Wess-Zumino-Witten-like $3\to 2$ process, relevant to the dark matter relic density generation mechanism~\cite{Hochberg:2014dra,Hochberg:2014kqa,Kulkarni:2022bvh}. There is also an additional $3\to 2$ process, already in the one-flavor theory, which converts dark pions into dark fermionic baryons, and allows to build Fermi pressure, from accrued bosonic dark matter, required for neutron star stabilization~\cite{Hajizadeh:2017jsw}. G$_2$-QCD therefore provides an ideal testbed for the previously mentioned challenge.

In this work, as a proof-of-principle, we use the available~\cite{Maas:2012wr,Wellegehausen:2013cya,Hajizadeh:2017jsw} equation of state for a dark G$_2$-QCD sector with one flavor\footnote{Since the chiral dynamics is the same for more flavors~\cite{Holland:2003jy,Wellegehausen:2013cya,Kogut:2000ek}, and up to multiplicities also the spectrum, we do not expect too many non-trivial changes to the equation of state when increasing the number of flavors.}, and build neutron stars made from both dark matter and ordinary matter. This equation of state has been obtained using lattice simulations~\cite{Maas:2012wr,Wellegehausen:2013cya}, and thus fully non-perturbatively. However, it lacks the quantitative precision of ordinary equations of state obtained indirectly, due to limitations in terms of computing time~\cite{Wellegehausen:2013cya}. Nonetheless, as already seen in the studies of pure dark objects~\cite{Hajizadeh:2017jsw} the accuracy should be more than sufficient for an exploratory study of the viability of such a scenario.

For ordinary matter, we use a selection of the currently favored equations of state~\cite{Kurkela:2014vha, Dengler:2021qcq, Pitz:2024xvh} covering a wide range of phenomenology, and as much as possible data-driven constraints. This allows us to minimize the model-dependency of our investigation. The results will therefore only be quantitatively relevant if either we keep the amount of dark matter limited such that, within errors, neutron star properties remain the same. Or we stipulate that most neutron stars do not contain large amounts of dark matter, but some do, giving a distinct signal differing from ordinary neutron stars when observed. Alternatively, one can interpret our injection of G$_2$-QCD equation of state seriously and demand modification of the ordinary matter equation of state to compensate for large modifications in the neutron star properties. However, this is beyond the scope of the present exploratory work, especially in view of the restricted set of available equations of state for the G$_2$-QCD case. We will also assume throughout that the interactions between dark matter and ordinary matter, even at large densities of both, are negligible. This may be justified given that in most strongly-interacting dark matter scenarios the portal couplings to the SM are highly suppressed~\cite{Hochberg:2015vrg, Berlin:2018tvf, Pomper:2024otb, Bernreuther:2023kcg}.

For the sake of completeness, we discuss briefly our choices of the equations of state in \Cref{s:eos}, together with a brief overview of the salient features of G$_2$-QCD. Our approach to building a two-component neutron star is standard~\cite{Hajkarim:2024ecp}, and briefly rehearsed in \Cref{s:ns}. In \Cref{s:res} we provide results for a range of dark matter masses and dark matter content. We provide the mass-radius relations as well was the tidal deformability, which is the primary input to the gravitational wave signal for the merging of two neutron stars. Finally, taking our inputs at face value, we interpret our result in terms of limits on dark matter properties or dark matter contents in neutron stars. This part should be considered rather a blueprint than a hard statement, given our limited inputs, but can be improved arbitrarily by improving these inputs. Additional results are available in \Cref{a:res}.

\section{Equations of state}\label{s:eos}

\begin{figure}
\begin{center}
    \includegraphics[width = 0.49\textwidth]{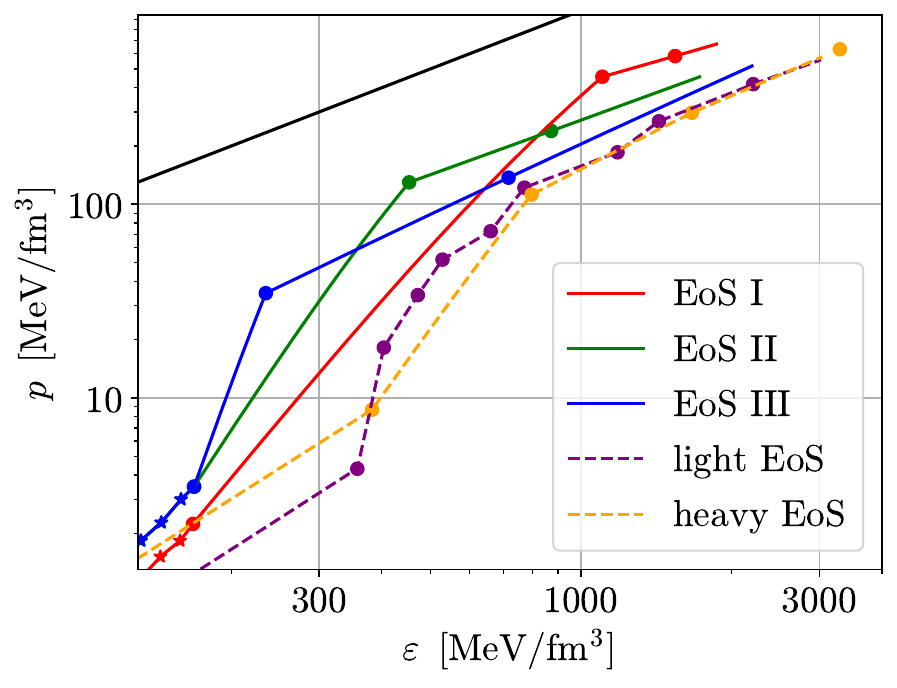}
    \includegraphics[width = 0.49\textwidth]{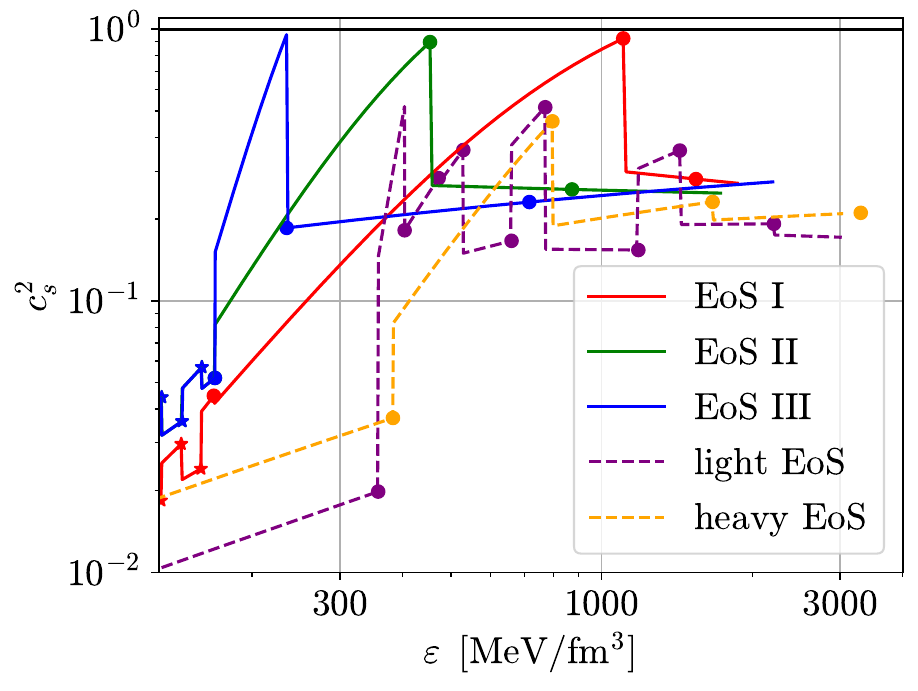}
    \caption{The equations of state (left) and the speed of sound (right) from~\cite{Kurkela:2014vha} and~\cite{Wellegehausen:2013cya} used in this work. The dark matter candidate mass is chosen to be the mass of the neutron for this plot for comparison. The end of the lines indicate the maximum value for the pressure for the respective equation of state used in this work. The black line indicates $\varepsilon = p$ and $c_s^2=1$ respectively. For the ordinary matter equation of state, the stars denote the results from NChPT while the dots show the endpoints of the piecewise polytropes. For the dark matter equation of state, the points indicate lattice data with polytropes, see appendix \Cref{a:interpolation}.}
    \label{plot:EoS}
\end{center}
\end{figure}

The equation of state (EoS) is the key ingredient for the Tolman-Oppenheimer-Volkoff (TOV) equation~\cite{Oppenheimer:1939ne} which is commonly used to calculate macroscopic neutron star observables. To describe a neutron star consisting of two fluids we need the EoS of both of them. A novelty of this work is the use of an EoS of a confining, asymptotically free gauge theory obtained from first principles. The latter is crucial for a strongly-interacting dark matter model, as only this will ensure non-trivial constraints between different properties, e.\ g.\ the masses of stable states, resonances, and the EoS. We want to complement the dark matter EoS thus, naturally, by the use of an EoS for ordinary matter in as much a model agnostic way that is able to meet the observational constraints. In this way we minimize the number of free parameters. We emphasize that this choice is not sufficient to cover all possible scenarios for neutron stars, as equations of states with (first order) phase transitions or twin-stars, for example, are not covered \cite{Mendes:2024hbn,Christian:2023hez}. 

Both, ordinary and dark matter, EoS will be needed to be interpolated from separate, partly discrete, information. This will be done using polytropes \cite{Kurkela:2009gj}. Details of this procedure are given in \Cref{a:interpolation}.

\subsection{Ordinary matter}

For ordinary matter we work with the EoS from~\cite{Kurkela:2014vha}. This work uses piecewise polytropes, to interpolate between a low density regime described by nuclear chiral perturbation theory (NChPT)~\cite{Hebeler_2013} and a high density regime described by perturbative QCD~\cite{Kurkela:2009gj}. Perturbative QCD is only valid at densities much larger than those found in neutron stars, but asymptotically will describe the EoS. Thus, following~\cite{Kurkela:2014vha}, an interpolation between the two regimes should be possible within limits like the speed of sound $c_s$ satisfying $c_s<1$. 

For this study, we use 3 limiting equations of state (EoS I, II and III) from~\cite{Kurkela:2014vha}. EoS I is the softest of the three, barely reaching the two solar mass limits. EoS II yields a maximal maximum mass. EoS III is the stiffest and while it might be too stiff to describe an ordinary neutron star, we still include it in our analysis as the addition of dark matter to a neutron star can result in smaller masses and radii. Therefore it might be able to meet observational constraints when considering mixed neutron stars. For the same reason we will employ central pressures larger than the ones stated in~\cite{Kurkela:2014vha} that result in the maximal mass. We will see that the addition of dark matter allows for stable solutions at larger central pressures  \cite{Kain:2021hpk,Valdez-Alvarado:2012rct}. \Cref{plot:EoS} shows the pressure and the speed of sound versus the energy density for the three ordinary matter and two dark matter equations of state employed.

We extend the EoS below the NChPT data points in~\cite{Kurkela:2014vha} by interpolating from zero using another polytrope with adiabatic index $\Gamma=5/3$. This corresponds to a free fermi gas and coincides with the equation of state $n\propto (\mu^2-m_n^2)^{3/2}$ for low densities. We do not include a specific EoS for the crust, which would affect the radius. This should be considered a systematic uncertainty, but likely a subleading one for the present purpose.

\subsection{G$_2$-QCD}

G$_2$-QCD is a QCD-like theory with the gauge group SU(3) replaced by the exceptional Lie group G$_2$ ~\cite{Holland:2003jy,Pepe:2006er}. We consider the situation with one flavor of fermions in the fundamental representation~\cite{Maas:2012wr,Wellegehausen:2013cya}, called dark quarks in the following. G$_2$-QCD has, due to its group structure, a number of extraordinary features. Most relevant here is that chiral symmetry breaking occurs already in the one-flavor theory, and that dark baryons can be built with any number of dark quarks~\cite{Kogut:2000ek,Holland:2003jy,Wellegehausen:2013cya}. It can also be simulated at all densities using first-principle lattice methods~\cite{Maas:2012wr} using standard algorithms~\cite{Gattringer:2010zz,Wellegehausen:2011sc}. This will necessarily capture all self-interaction effects on the EoS completely.

A detailed discussion of its spectrum and other properties in the vacuum can be found in ~\cite{Holland:2003jy,Pepe:2006er,Wellegehausen:2013cya}. The most relevant features for our purpose here is that the lightest absolutely stable fermionic state is supposedly a three dark-quark state. In addition, there can exist a one dark-quark hadron ~\cite{Holland:2003jy}, but it is likely heavier for the dark quark masses employed here, because it is a dark-quark-dark-gluon hybrid state~\cite{Hajizadeh:2017jsw,Wellegehausen:2013cya}. It is thus supposedly unstable against decay into the three dark-quark state and dark mesons. Note that in G$_2$-QCD dark quarks and dark antiquarks, as e.\ g.\ also in Sp(4)-QCD and SU(2)-QCD, cannot be distinguished~\cite{Kogut:2000ek}, and thus only Grassmann parity is conserved. In fact, the dark baryonic chemical potential is a Weyl-flavor potential, but serves the same purpose as the baryonic chemical potential in QCD~\cite{Maas:2012wr,Wellegehausen:2013cya}. As a consequence, dark mesons cannot be distinguished from dark glueballs in the one-flavor theory, and thus are only stable if the theory remains in isolation. Hence, we identify the three dark-quark state as the dark matter candidate of our theory. We therefore use its mass also for scale setting. At the current time, low-energy features beyond the spectrum in \cite{Wellegehausen:2013cya} have not been obtained. In particular, establishing a low-energy effective theory in the spirit of \cite{Kulkarni:2022bvh} has not (yet) been done, and would require a major effort. However, the low-energy effective theory is never explicitly needed here, and all non-trivial effects on the EoS are automatically cpatured by the lattice simulations.

Details on the EoS, its determination and properties, can be found in ~\cite{Wellegehausen:2013cya,Maas:2012wr,Hajizadeh:2017jsw}. Here, we concentrate on the EoS at zero temperature, which has been discussed in~\cite{Wellegehausen:2013cya,Hajizadeh:2017jsw}. Of course, the lattice data is only available at discrete values of the dark baryonic chemical potential. The interpolation between data points is, as for the ordinary matter equation of state, performed using piecewise polytropes to ensure thermodynamic consistency.

Just like for the ordinary matter EoS we use choose $\Gamma=5/3$ for the polytrope describing the low density behavior. This coincides with $n=c(\mu^2 - m_C^2)^{3/2}$ for low densities, where $m_C$ is the mass of the dark matter candidate and c is a free parameter as was considered in \cite{Hajizadeh:2017jsw}. For the light ensemble and at high densities, we extrapolate the data using the fit form $n(\mu) = n_s/(\exp(a-b \mu)+1)$ fitted to the last few data points~\cite{Wellegehausen:2013cya,Hajizadeh:2017jsw}. Herein $n_s$ is the lattice saturation density, i.\ e.\ the situation where Pauli blocking obstructs larger densities. This is a pure lattice artifact but the densities required in the following, as has already been observed previously~\cite{Hajizadeh:2017jsw}, are small enough for this to be not a relevant issue. Indeed, for the heavy ensemble below, the extrapolation region is never probed. The interpolation and extrapolation captures the main physical features of the lattice results while providing access to a larger range in the density. Note that, in contrast to the low-density case, the high-density fit function has not been chosen on physical grounds, but for its ability to describe the lattice data very accurately.

The EoS has been obtained for two different values of the bare lattice parameters\footnote{After scale setting, i.\ e.\ choice of the dark matter mass in GeV, the only free parameter in the theory for the dark matter is the bare fermion mass.}, which will result in different mass spectra. The lightest particle in the theory is the pseudo-scalar \textit{dark pion} which is unstable to potential decays to the standard model via messengers. The dark matter candidate is the lightest fermionic particle. In the two resulting ensembles the mass of the dark pion is 0.25$\,m_c$ and 0.33$\,m_c$ respectively. The original choice of the lattice simulation parameters were for a different reasons than dark matter \cite{Maas:2012wr}, and thus do not apply here. It would require a major effort and incur computational costs of the order of a few tens of millions of core hours to provide further values~\cite{Wellegehausen:2013cya}. This would only be justified by concrete observational hints for a particular mass value. However, for our exploratory investigations, the precise value does not matter too much. As will be seen, the impact is only a limited quantitative one, and the change between the ordinary matter EoS has already a more substantial effect. The resulting dark matter EoS are also shown in \Cref{plot:EoS}.

\section{Building a two-component neutron star}\label{s:ns}

In this section we will describe the framework used to build a neutron star from the two-fluid TOV equation~\cite{Tolos:2015qra,Hajkarim:2024ecp}. For comparison, we depict the different involved EoS in \Cref{plot:EoS}.
\subsection{Two-fluid TOV}\label{s:tov}
The TOV equation is a system of coupled differential equations describing the pressure gradient $dp/dr$ and mass gradient $dm/dr$ derived from the conservation of the energy stress tensor and is consistent with general relativistic effects. Assuming a static, spherically symmetric object in hydrostatic equilibrium, one obtains the ordinary TOV equation consisting of the pressure and a mass gradient. While usually used for determining ordinary matter properties, a set of two-fluid TOV equations containing an exotic matter component can be achieved by neglecting the interactions between ordinary and exotic matter components. The following set of equations are the two-fluid TOV equations written in a dimensionless form~\cite{Narain:2006kx,Tolos:2015qra}. In particular, after switching to natural units, we rescale all dimensionful quantities in units of the mass of the dark matter candidate $m_{C}$, yielding
\begin{align}
\begin{split}
\frac{dp_{O}}{dr} &= -(p_{O} + \varepsilon_{O} ) \frac{d\nu}{dr}  \\
\frac{dm_{O}}{dr} &= 4\pi r^2 \varepsilon_{O} \\
\frac{dp_{D}}{dr} &= -(p_{D} + \varepsilon_{D} ) \frac{d\nu}{dr}  \\
\frac{dm_{D}}{dr} &= 4\pi r^2 \varepsilon_{D}
\end{split}
\end{align}
where the subscript $O$ refers to the ordinary matter fluid\footnote{Sometimes this is also called baryonic matter. We decide to use ordinary ($O$) as our dark matter is also (dark) baryonic.} while $D$ stands for dark matter fluid. $\varepsilon$ is the energy density. In the following, lowercase $m$ and $r$ will refer to the mass and radius during the integration. Capital letters $M_{D/O}$ and $R_{D/O}$ refer to the neutron star mass and radius, respectively, where subscripts $D$ and $O$ denote the corresponding total dark and ordinary values for these quantities, respectively. Finally, the metric function $d\nu/dr$ is defined for both fluids as
\begin{equation}
\frac{d\nu}{dr}= \frac{(m_{O}+m_{D})+4\pi r^3(p_{O}+p_{D})}{r\left(r-2(m_{O}+m_{D})\right)}.
\end{equation}

\noindent The coupled two-fluid TOV equations can be solved using the EoS of the two fluids. Iterating over the potentially nonidentical central pressures, one obtains the neutron star properties for each combination of central pressures ($p_{0,O}$ and $p_{0,D}$) by integrating the equations until the pressure vanishes. The radius at which the ordinary/dark matter pressure drops to zero within numerical precision defines $R_{O}$/$R_{D}$. $M_O$/$M_D$ are the resulting integrated masses of the two fluids respectively and $M_{tot}=M_{O}+M_{D}$. The solutions can result in a neutron star with a dark core ($R_O>R_D$) or a dark halo ($R_O<R_D$), which potentially also affects the gravitational wave signal of an inspiral of two neutron stars~\cite{Ivanytskyi:2019wxd}.

Dimensional analysis shows that the mass and the radius scale like $1/m_N^2$. Given an equation of state that can be trivially scaled by the mass of its constituents $m_{const}$, as is the case with our dark matter EoS, halving the mass of the constituents would result in a star with four times the mass and four times the radius. To some extent, this scaling argument is still present in the two fluid case. A lighter dark matter candidate mass will result in a stronger impact of dark matter on the star. In the limiting cases where the central pressure of one of either the dark or ordinary matter is negligible ($p_{0,D} \gg p_{0,O}$ or $p_{0,D} \ll p_{0,O}$) this scaling argument is perfectly recovered as very heavy dark matter candidates will have no impact on the observables while a light dark matter candidate will result in a compact object dominated by dark matter that can eventually not be interpreted as a neutron star anymore.
\subsection{Stability}\label{ss:stability}
Not every solution of the TOV equations yields a stable neutron star. Small perturbations in the metric field should settle back to the original solution. By solving a Sturm-Liouville problem one can identify the stable solutions by demanding the following criterion~\cite{Hippert:2022snq,Pitz:2024xvh,Barbat:2024yvi}
\begin{equation}\label{eq:stability}
    \begin{pmatrix}
\delta N_{O} \\
\delta N_{D}
\end{pmatrix}=
\begin{pmatrix}
\partial N_{O}/\partial \varepsilon_{c,O} & \partial N_{O}/\partial \varepsilon_{c,D} \\
\partial N_{D}/\partial \varepsilon_{c,O} & \partial N_{D}/\partial \varepsilon_{c,D}
\end{pmatrix}\begin{pmatrix}
\delta \varepsilon_{c,O} \\
\delta \varepsilon_{c,D}
\end{pmatrix}=0,
\end{equation}

\noindent where $N_{O/D}$ is the total number of ordinary/dark particles and $\varepsilon_{c,O/D}$ is the energy density of ordinary/dark matter in the center of the star. This problem can be translated into demanding that both the eigenvalues of the equation in \eqref{eq:stability} be positive. The number of particles can be calculated simultaneously to the TOV equations via
\begin{equation}\label{eq:dN/dr}
    dN = 4 \pi \left( 1-\frac{2m}{r}\right)^{-1/2}n r^2 dr.
\end{equation}

\noindent For polytropes, the number density $n$ is given by

\begin{equation}
    n = \left( \frac{p}{K}\right)^{1/\Gamma},
\end{equation}
\noindent where, K is a constant and $\Gamma$ depends on the polytropic index n like $\Gamma=\frac{\text{n}+1}{\text{n}}$~\cite{Sagert:2005fw}.
\subsection{Tidal Deformability} \label{s:tidal}
When two compact objects orbit each other, they get deformed by the gravitational tidal field of one another. The tidal deformability $\Lambda$ is a quantity that describes the proportionality between the tidal field and the induced quadrupole deformation. We will work with the dimensionless version of the tidal deformability that is connected to the dimensionful tidal deformability $\lambda$ and the dimensionless second Love number $k_2$ via 

\begin{equation}
\Lambda = \frac{\lambda}{M_{tot}^5}=\frac{2}{3}\frac{k_2}{C^5}.\label{tidal}
\end{equation}
Here, we have introduced the compactness $C=M_{tot}/R$ calculated with the total mass $M_{tot}=M_O + M_D$ and the maximal radius $R=\max(R_O, R_D)$. Calculating the tidal deformability requires the second Love number, which can be calculated from the neutron star compactness as
\begin{align}\label{k_2_from_y}
\begin{split}
k_2 =
&\frac{8C^5}{5}
 \left(1-2C\right)^2 \left[2+2C(y-1)-y\right]\times\\ 
&\left\{2C\left[6-3y+3C(5y-8)\right]+\right.\\
&4C^3\left[13-11y+C(3y-2)+2C^2(1+y)\right]+\\
&\left. 3\left(1-2C\right)^2\left[2-y+2C(y-1)\right]\ln(1-2C)\right\}^{-1},
\end{split}
\end{align}
\noindent The auxiliary parameter $y$ can be integrated simultaneously to the TOV equation by
\begin{equation}\label{dy_dr}
r\frac{dy(r)}{dr} + y(r)^2+y(r)F(r) + r^2Q(r) =0
\end{equation}
where 
\begin{equation}
    F(r) = \frac{r-4\pi r^3 ((\varepsilon_{O}+\epsilon_{D}) - (p_{B}+p_{D}))}{r-2m(r)}
\end{equation}
and
\begin{align}\label{Q_r}
\begin{split}
    Q(r) =& \frac{4\pi r\left(5(\varepsilon_{O}+\varepsilon_{D})+9(p_{O}+p_{D})+\frac{\varepsilon_{O}+p_{O}}{c_{s,O}^2}+\frac{\varepsilon_{D}+p_{D}}{c_{s,D}^2}-\frac{6}{4\pi
 r^2}\right)}{r-2m(r)}\\
&-4\left[\frac{m(r)+4\pi r^3 (p_{O} + p_{D})}{r^2(1-\frac{2m(r)}{r})}\right]^2
\end{split}
\end{align}
with $m=m_{O}+m_{D}$ and $c_{s,O/D}^2 = dP_{O/D}/d\varepsilon_{O/D}$ is the speed of sound~\cite{Barbat:2024yvi}. 

In contrast to black holes, neutron stars have a non-vanishing tidal deformability. As a result, the gravitational wave signal of a neutron star binary differs from the one produced by a black hole merger by a (potentially time-dependent) phase in the late stages of the inspiral. To leading order in the tidal deformabilities of the two neutron stars, this phase can be determined by the parameter $\tilde{\Lambda}$ as~\cite{LIGO_GW170817}
\begin{equation}
    \tilde{\Lambda} = \frac{16}{13} \frac{(M_1+12M_2)M_1^4\Lambda_1+(M_2+12M_1)M_2^4\Lambda_2}{(M_1+M_2)^5}
\end{equation}
\noindent where $\Lambda_1$ and $\Lambda_2$ are the tidal deformabilities and $M_1$ and $M_2$ are the total masses of the two neutron stars. This definition is chosen such that $\tilde{\Lambda}=\Lambda_1=\Lambda_2$ if $M_1=M_2$. Therefore, we can use it to constrain our results at the mass extracted from the merger event. We use the low-spin prior results for which this quantity is constraint to <800 for $M_\text{tot}=1.38^{+0.24}_{-0.19}M_\odot$ from GW170817~\cite{LIGOScientific:2017vwq,LIGOScientific:2017ync} and <600 for $M_\text{tot}=1.6^{+0.27}_{-0.14}M_\odot$ from GW190425~\cite{LIGOScientific:2020aai}. Deviations in this quantity compared to an ordinary matter neutron star give a measure of the impact a dark matter component will have on the gravitational wave signal of two compact stellar objects. 

The applicability of gravitational wave analysis beyond black-hole signals to more complex and exotic systems such as the mixed stars studied here is not fully understood~\cite{Cardoso:2016rao,Cardoso:2017cqb,DiGiovanni:2022mkn,Bauswein:2020kor}. As the nature of this work is exploratory, we will discuss the results of the tidal deformability for all our results for illustrative purposes only.
\section{Results}\label{s:res}
In this section, we show our results for the TOV equation. We investigated the three EoS of state for ordinary matter as well as two EoS for dark matter described by one-flavor G$_2$-QCD at two different values of the bare quark mass labeled as \textit{light} and \textit{heavy}. They correspond to the lightest particle, the dark pion, having a mass of 25\% and 33\% of our dark matter candidate, respectively~\cite{Wellegehausen:2013cya}. We will discuss the main findings by showing the results for EoS II and the light dark matter equation of state. The results for the remaining combinations of equations of state can be found in \Cref{a:res}, and differ only quantitatively. On top of that, we investigate the mass $m_C$ of the dark matter candidate from 500 MeV to 4 GeV, and show four benchmark values for $m_C$ in the figures. This range is well motivated by other studies on self-interacting dark matter and covers all interesting dynamics in the system~\cite{Kulkarni:2022bvh,Dengler:2024maq,Hochberg:2014kqa}. Also, fixing the scale is necessary as the combination of Newton's constant and the mass of the neutron in ordinary matter sets an absolute scale.

For dark matter masses much smaller or larger than the neutron the solutions can be obtained by scaling arguments. For a single fluid compact object whose EoS can be trivially scaled by some scale (the dark matter candidate mass in the case of our dark matter EoS), the mass and radius scale like $m^{-2}$ where $m$ is the scale. This scaling argument is recovered for very light dark matter, as it starts to dominate the neutron star. For heavy dark matter, we obtain an ordinary neutron star dominated by standard model matter. We expect a balance between the impact of the two fluids at similar mass scales, as, despite all differences in details, the overall scales of G$_2$-QCD are still similar to ordinary QCD~\cite{Wellegehausen:2013cya}.


\begin{figure}[h!]
\begin{center}
    \includegraphics[width = 0.95\textwidth]{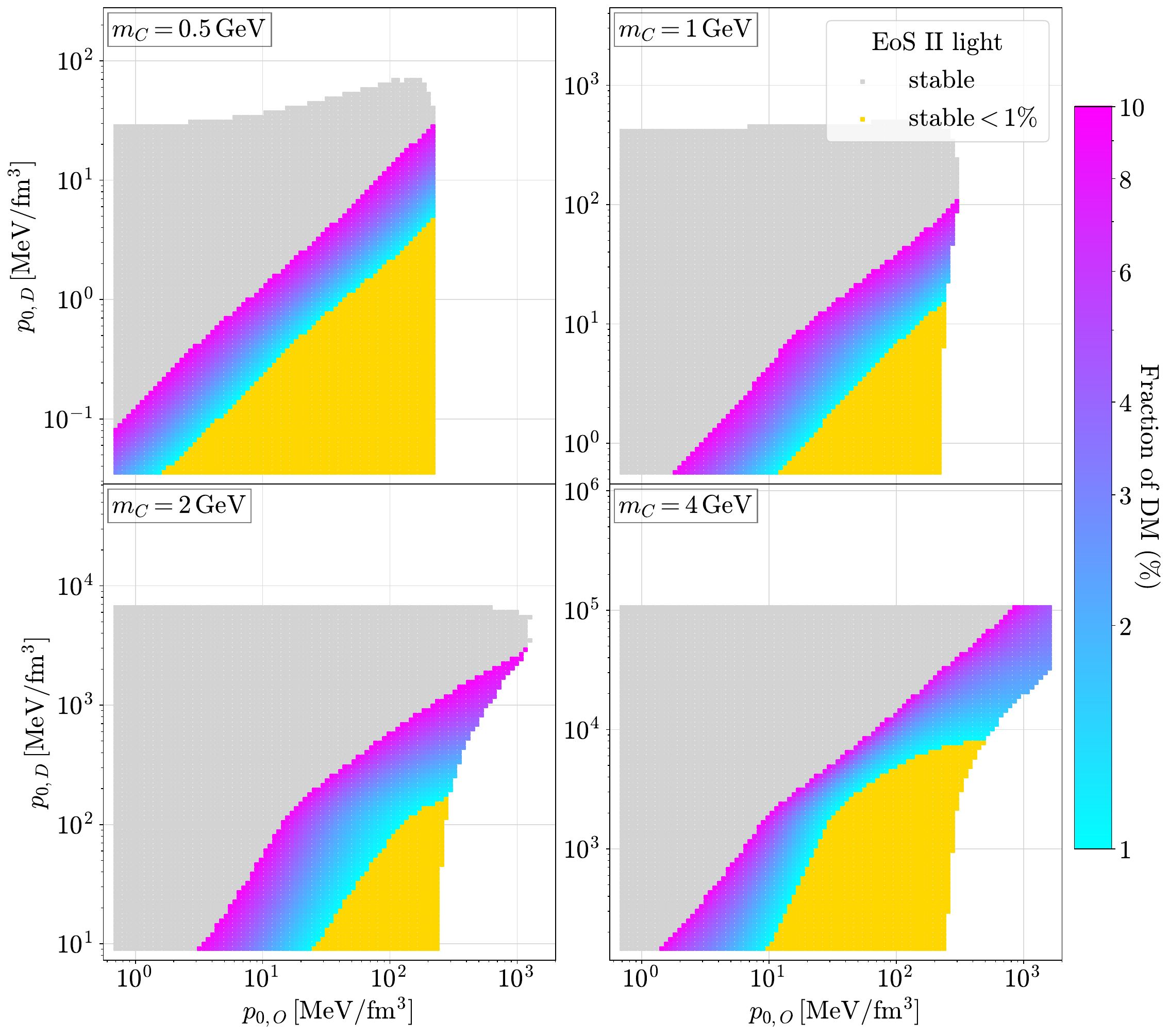}
    \caption{The investigated set of central pressures of ordinary matter ($p_{0,O}$) and dark matter ($p_{0,D}$) for EoS II and the light dark matter equation of state. We will use this combination of ensembles in the main text and show further results in \Cref{a:res}. Gray points indicate stable solutions with a dark matter fraction >10\%. Colored points indicate the dark matter fraction and we indicate solutions with dark matter fractions $<$1\% in yellow. The same color scheme is used in the following figures. We show four representative dark matter candidate masses from 0.5 GeV to 4 GeV.}
    \label{plot:pressure}
\end{center}
\end{figure}

We solve the two-fluid TOV equation by iterating over both the central pressures of ordinary matter and dark matter and perform the stability analysis. In \Cref{plot:pressure} we show the range of central pressures that we investigated for EoS II and the light EoS for dark matter. The further results depend on the amount of dark matter in the object.

Different publications use different realistic upper bounds for the amount of dark matter in neutron stars~\cite{Ivanytskyi:2019wxd,Miao:2022rqj,Giangrandi:2022wht,Rutherford:2022xeb}. Consequently, we will provide results not for a fixed amount of dark matter, but show the impact of a varying amount.  In the figures, gray points indicate a stable solution to the TOV equation, but with a dark matter component larger than 10\%. The further color coding indicates the dark matter fraction $M_{D}/M_\text{tot}$ below 10\%. In particular, yellow points indicate solutions with dark matter fractions <1\%. With our choice of 1\% we want to ensure that the resulting compact objects can actually be identified with neutron stars. In general, for a dark matter component below 1\% of the total mass, the results are more strongly affected by the choice of the ordinary matter EoS employed and than by the mass of the dark matter candidate.

One immediate observation is that the addition of (light) heavy dark matter makes it possible to sustain larger central pressures for (dark) ordinary matter as was also found in \cite{Kain:2021hpk,Valdez-Alvarado:2012rct}. This is particularly interesting as this grants access to new regimes of EoS that could potentially have interesting phenomenology. We see that in the 4\,GeV case, even when the dark matter fraction is kept at <1\%, larger central pressures of ordinary matter are possible.  We also find stable solutions that exceed this bound which could be labeled as non-standard compact stellar objects or dark stars. As noted in the introduction, if the latter would be the dominating population, our reliance on a fixed QCD EoS may become dubious.

\begin{figure}
\begin{center}
    \includegraphics[width = 0.95\textwidth]{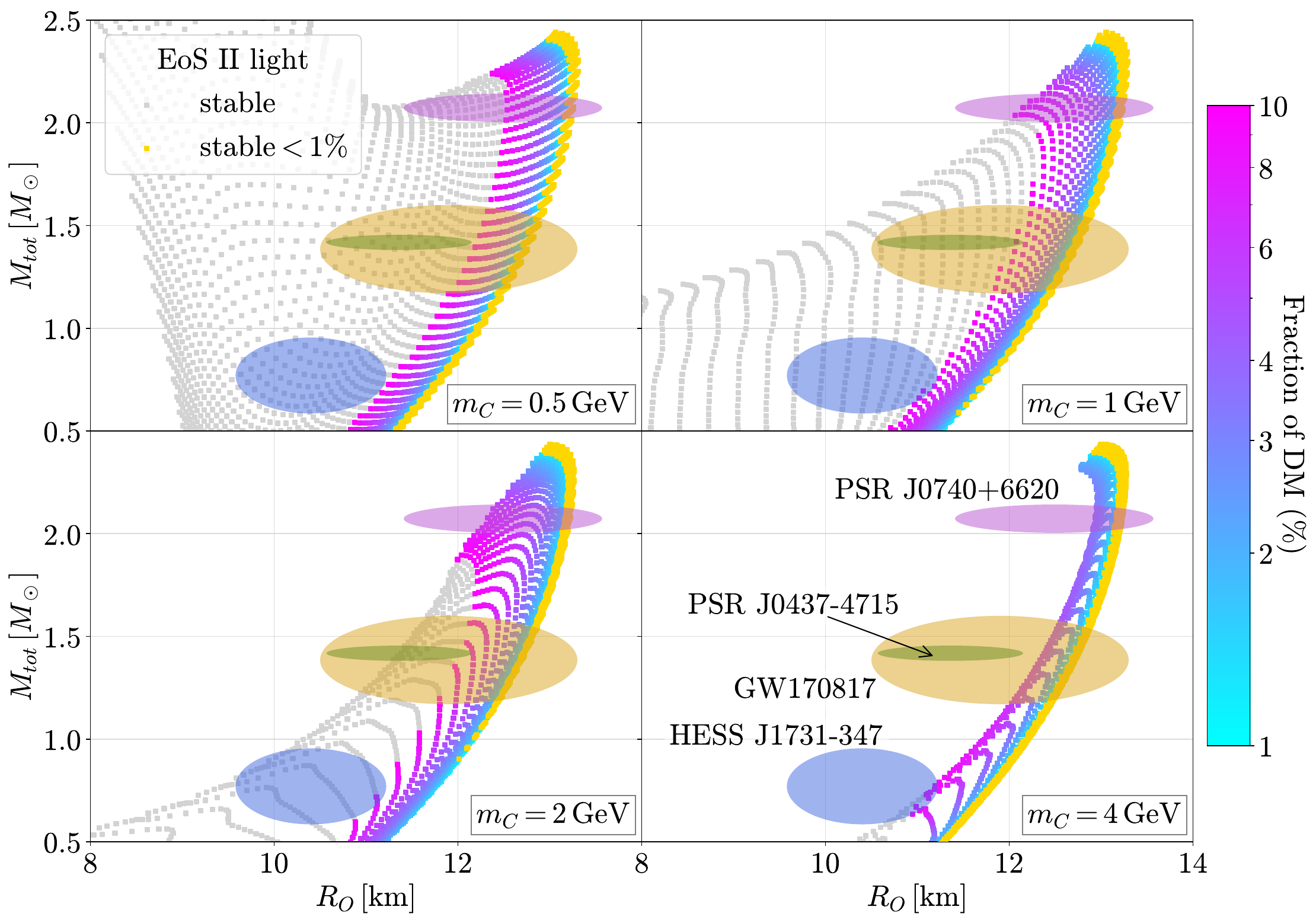}
    \caption{The mass-radius relation between the total observable mass $M_O+M_D$ in solar masses and the observable ordinary radius $R_O$ in km in the same color-coding and for the same equations of state as in \Cref{plot:pressure}. The ellipses show results for the mass and radius of various neutron stars, which were determined using different techniques~\cite{Doroshenko:2022nwp,Salmi:2024aum,Choudhury:2024xbk,LIGO_GW170817}.}
    \label{plot:mr}
\end{center}
\end{figure}

In \Cref{plot:mr} we show the mass-radius relation for all stable neutron stars using EoS II and the light equation of state. The y-axis is chosen to be the total mass, as mass measurements rely on gravitational signals. Radius measurements on the other hand, usually rely on electromagnetic observation, which is not sensitive to the dark matter substar. Therefore, the x-axis shows the ordinary matter radius. However, from figure \Cref{plot:mm_and_rr} below, it is visible that for sufficiently heavy dark matter all configurations we investigated with a dark matter fraction >10\% contain dark cores, where the ordinary matter radius is identical to the outermost radius. We consistently observe that the mass decreases with the addition of dark matter. This is because the additional dark matter results in a larger central pressure which in turn increases the value of the metric function and with that a larger pressure gradient for both ordinary and dark matter. This results in a smaller core of the star, which is the densest part and therefore contributes the most to the mass. For all dark matter candidate masses, we find decent agreements with the observation of neutron stars, albeit that this statement mostly depends on the ordinary matter equation of state employed. However, we see that at 1\% dark matter, the mass can shift by $\sim0.1\,M_\odot$ and the radius by $\sim100\,m$. For up to 10\% dark matter, the mass can decrease by $\sim0.4\,M_\odot$ and the radius can decrease by $\sim800\,m$. The apparent observation that \cite{Doroshenko:2022nwp,Salmi:2024aum,Choudhury:2024xbk,LIGO_GW170817} are better described with a substantial dark matter content than with our pure ordinary matter EoS should not be interpreted too much. We do not model the ordinary matter crust, which especially could affect the apparent radius, and we do not fully sample all proposed ordinary matter EoS. What should be taken from these results, however, is that added SIMP dark matter can easily hide inside the observations. Especially, if, as is the case presently, it shows a similar, large, stiffness as required by the observations. This is an inherent feature of our dark matter EoS, and a direct consequence of the details of its self-interaction captured by the lattice simulations.


\begin{figure}
\begin{center}
    \includegraphics[width = 0.49\textwidth]{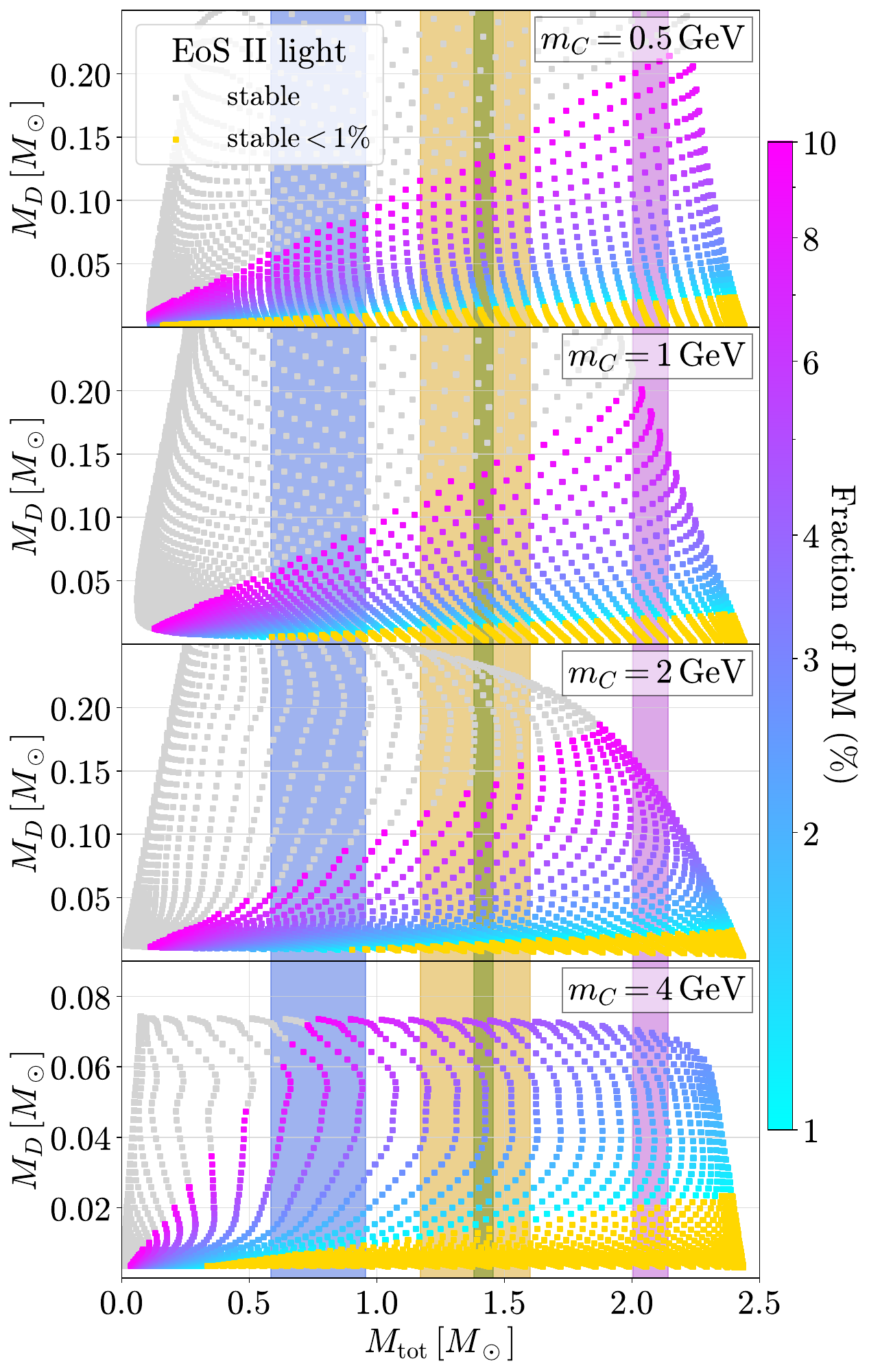}
    \includegraphics[width = 0.49\textwidth]{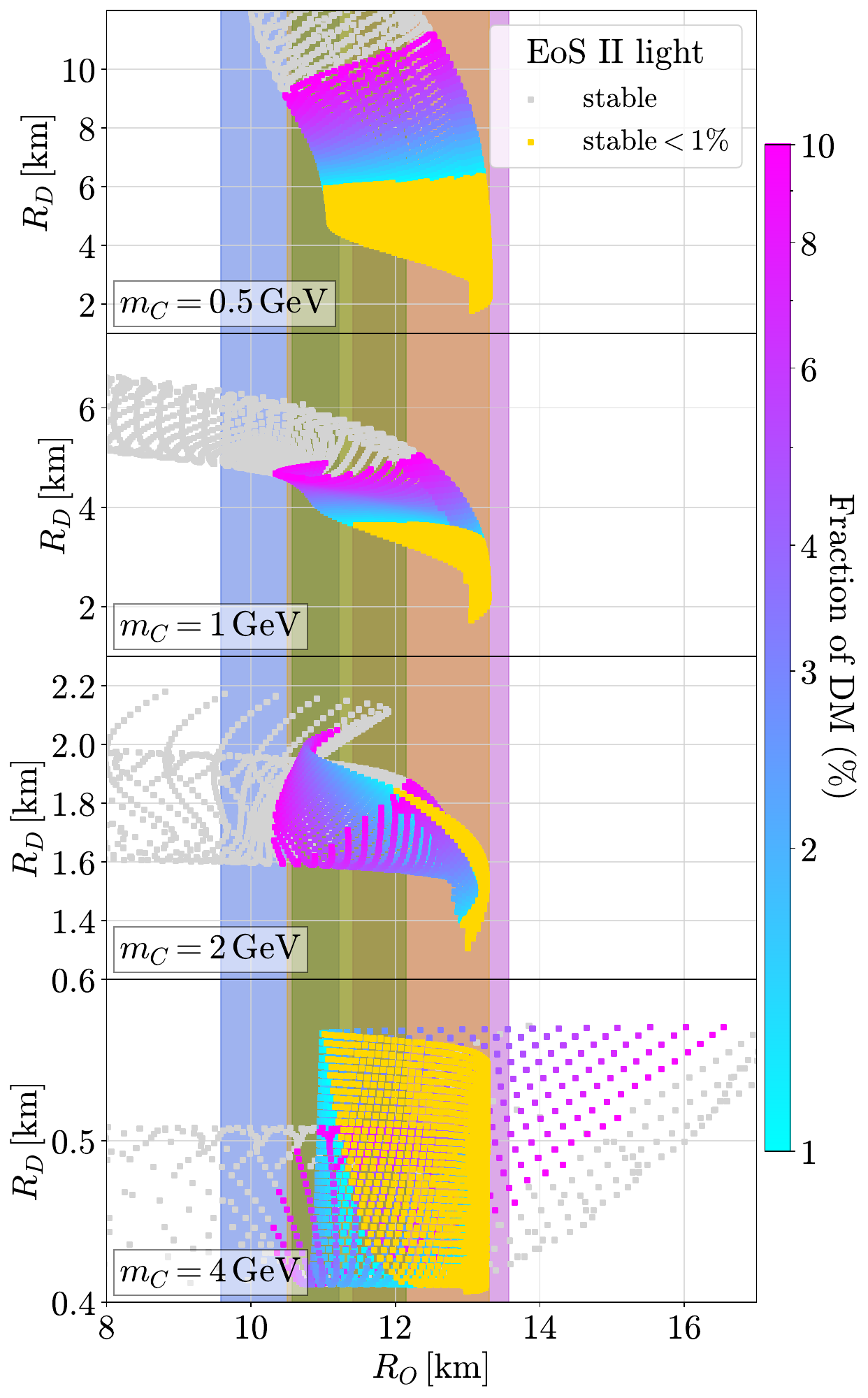}
    \caption{Left: The dark matter mass versus the total mass. Right: The dark matter radius versus the ordinary matter radius. The x-axes are chosen such that we can show the neutron star measurements from~\cite{Doroshenko:2022nwp,Salmi:2024aum,Choudhury:2024xbk,LIGO_GW170817}. Note the different y-axes for $R_D$.}
    \label{plot:mm_and_rr}
\end{center}
\end{figure}

\begin{figure}
\begin{center}
    \includegraphics[width = 0.95\textwidth]{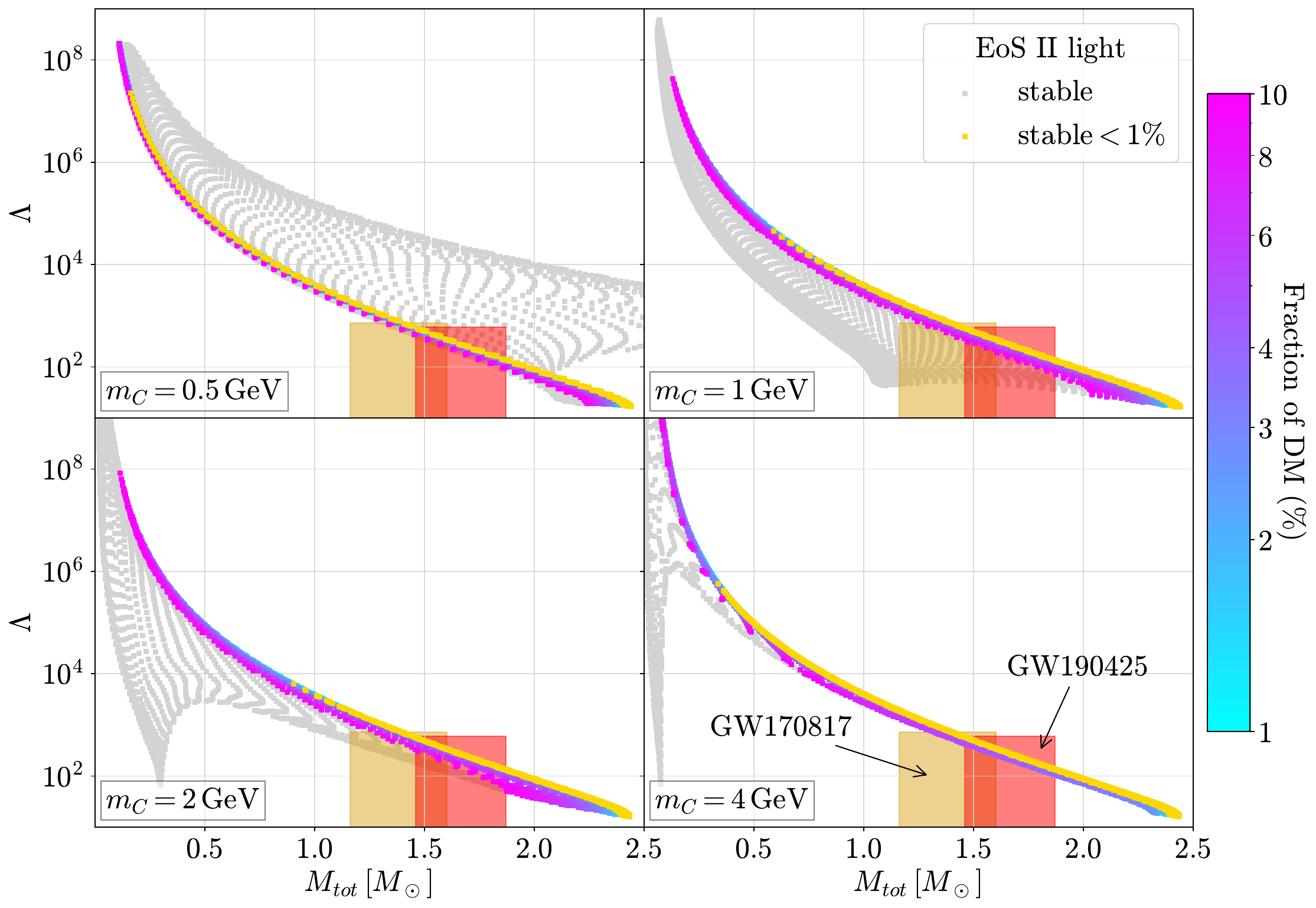} 
    \caption{The dimensionless tidal deformability versus the total mass. The yellow and red boxed are gravitational wave observation from \cite{LIGO_GW170817} and \cite{LIGOScientific:2020aai}.}
    \label{plot:lambda}
\end{center}
\end{figure}

In \Cref{plot:mm_and_rr}, we show the dark matter mass versus the total mass and the dark matter radius versus the ordinary matter radius. The x-axes are chosen such that we can show neutron star measurements as horizontal bands. As we restrict ourselves to 10\% dark matter in these plots, the total mass is mostly ordinary matter. In general, a light dark matter candidate will result in a larger dark matter extent. A heavy dark matter candidate on the other hand will result in a compact dark matter core. While the halo contributes more equally to the pressure gradient at all distances from the center of the star, the core has essentially no impact on the outer regions of the star. As a result, in the top panel on the left we see consistently an increase of the dark matter mass while the total mass decreases when the central pressure of dark matter is increased. In the bottom panel on the left, we see that increasing the dark matter central pressure first results in a larger total and dark matter mass. At some point, the dark core becomes large enough to influence the ordinary matter and results in a decreasing total mass at an increasing dark matter mass. On the right side, we see that the dark matter radius is much larger for small dark matter candidate masses. All configurations with a dark matter fraction <10\% are core configurations meaning that $R_D<R_O$. For a 0.5\,GeV dark matter candidate, the dark matter radius increases with increasing dark matter central pressure while the ordinary matter radius stays constant at first but decreases when the dark matter fraction exceeds 1\%. For 4\,GeV, we find that the dark matter radius increases only slightly by increasing the dark matter central pressure. The ordinary matter radius however decreases at first and increases afterwards roughly when the dark matter fraction exceeds 1\%.

Finally, we discuss the results of dimensionless tidal deformability versus the total mass shown for the same configurations and the same color scheme in \Cref{plot:lambda}. Gravitational wave detection could be more promising for the detection of dark matter in neutron stars because it is also sensible to dynamical information. An extended halo will alter the gravitational wave signal in a different way from a dark matter core, and the latter could result in a supplementary peak in the gravitational wave spectrum. This has been shown in simulations of binary neutron star events~\cite{Giangrandi:2025rko}. Our results cannot provide this dynamical information, but we can test our results with the constraints from neutron star merger observation. We find that when kept at <10\%, the mixed stars are compatible with the measurements. In general, the addition of dark matter will decrease the tidal deformability but is barely visible on the logarithmic scale. We see that the extended halos for a 0.5\,GeV dark matter candidate with dark matter fractions >10\% will result in much larger tidal deformabilities, which is expected given the scaling with $R^5$. Mixed stars at larger dark matter candidate masses will decrease the tidal deformability.

\section{Summary}\label{s:sum}

In this work we studied the impact of a strongly-interacting dark matter component on neutron star properties, using G$_2$-QCD. The EoS for the dark fluid was obtained at finite density from first principles using lattice field theory~\cite{Maas:2012wr,Wellegehausen:2013cya,Hajizadeh:2017jsw}. We used the two EoS at our disposal, differing in their mass scales, and found that they have similar influence on the resulting properties of the mixed star. On the standard model side, we covered a range of equations of state that capture the current constraints on the mass, radius, and tidal deformability. They were used in~\cite{Kurkela:2014vha} by interpolating between low and high densities using piecewise polytropes. 

As is well known from investigations of the ordinary matter EoS, they all result in different mass-radius-relations. The added dark matter has similar effects on all three of them. For each combination of EoS we varied the mass of the dark matter candidate and the central pressure of the ordinary fluid and dark fluid and performed a stability analysis. The addition of dark matter, in general, allows for larger possible central pressures which should be kept in mind when investigating ordinary matter EoS \cite{Kain:2021hpk,Valdez-Alvarado:2012rct}. As expected, we find that light dark matter has a stronger effect on the properties of the neutron star and that by tuning the ratio of central pressures one is also able to find solutions that fulfill the observational bounds. For some combinations of the dark matter candidate mass and the ratio of central pressures we find mixed stars that do not resemble neutron stars but could provide an explanation for non-standard compact stellar objects, if encountered. An astrophysical observation of exotic compact objects also in gravitational wave experiments would motivate a more careful analysis of these objects. We leave this for future exploration.

It is noteworthy that, for the first time, this investigation used an EoS from first principles for describing strongly-interacting dark matter in a two component neutron star. In thus takes the impact of any self-interaction on the EoS fully and consistently into account, yielding e.\ g.\ its large stiffness uniquely. This is motivated by the class of SIMP dark matter models. While the impact on, e.\ g., the mass-radius-relation may be similar to using an effective model, first principle calculations of a UV-complete confining theory that is also capable of addressing small-scale structure problems are an important step towards unveiling the true nature of dark matter.

\section*{Acknowledgments}

We are grateful to J.~Schaffner-Bielich and V.~Sagun for helpful discussions. Y.~D. has been supported by the Austrian Science Fund research teams grant STRONG-DM (FG1). S.~K. is supported by the Austrian Science Fund research teams grant STRONG-DM (FG1) and project number P 36947-N. S.~K. thanks the University of W\"urzburg for hospitality through the RTG 2994 program, while part of this work was completed.
\appendix
\section{Appendix}
\subsection{Method of interpolation using polytropes}\label{a:interpolation}

\begin{figure}
    \centering
    \includegraphics[width=0.7\linewidth]{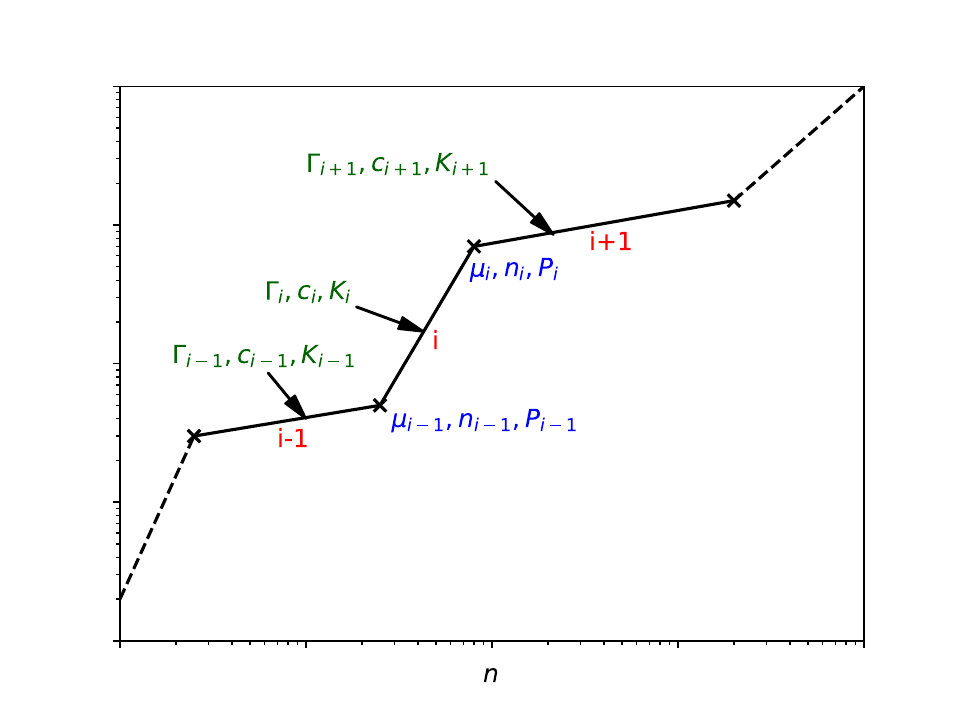}
    \caption{Labelling of the parameters of the piecewise polytropes.}
    \label{fig:piecewise_labelling}
\end{figure}

In order to use the lattice results for a calculation of the TOV equation we have to convert them to pressure $p$ and energy density $\varepsilon$. In order to ensure thermodynamic consistency at the level of guaranteeing $n\mu = p + \varepsilon$, we use the framework of piecewise polytropes to interpolate between the data points

Following \cite{Read:2008iy}, the polytropic equation of state is defined by \cite{Kurkela:2014vha}

\begin{equation}
    p=K n^\Gamma
\end{equation}

\noindent where $n$ is the number density and $\Gamma$ is called the adiabatic index. The energy density is fixed by

\begin{align}
\begin{split}
    d\left(\frac{\varepsilon}{n}\right) &= -p\,\,d \left(\frac{1}{n}\right)\\
    \varepsilon &= c n + \frac{K}{(\Gamma-1)}n^\Gamma ,
\end{split}
\end{align}

\noindent where in the second line an integration has been performed, introducing an integration constant $c$. From $n\mu = p + \varepsilon$ follows for the chemical potential.

\begin{equation}
    \mu = c + \frac{\Gamma}{\Gamma-1}Kn^{\Gamma-1}
\end{equation}

\noindent This formalism is used to interpolate between data points, see \Cref{fig:piecewise_labelling} for an illustration. The $i$'th piecewise polytrope, valid from $n_{i-1}$ to $n_{i}$, depends on the parameters $c_i$, $K_i$ and $\Gamma_i$. In the following, we will denote the functions depending on the number density and the $i$'th parameters as $p_i(n)$, $\varepsilon_i(n)$ and $\mu_i(n)$, and the values of these functions at $n_i$ as $p_i$, $\varepsilon_i$, $\mu_i$. 

At low-density follows from $\lim_{n\to0} \mu = m_c$ that for the first polytrope, $c_0=m_c$. The other assumption is that at very low density $\Gamma_0=\frac{5}{3}$ holds, which describes a free Fermi gas. This parameterization coincides with $n=c(\mu^2-m_c^2)^{\frac{3}{2}}$ at low densities. Demanding continuity, one can use the following formulas to find the parameters of the next piece of the polytrope,

\begin{align}
\begin{split}
    \mu_{i+1}-\mu_{i} &= \mu_{i+1}(n_{i+1})-\mu_1(n_{i}) \\
    &= c_{i+1}+K_{i+1}\frac{\Gamma_{i+1}}{\Gamma_{i+1}-1}n_{i+1}^{\Gamma_{i+1}-1}-c_{i+1}-K_{i+1}\frac{\Gamma_{i+1}}{\Gamma_{i+1}-1}n_{i}^{\Gamma_{i+1}-1} \\
    &= K_{i+1} \frac{\Gamma_{i+1}}{\Gamma_{i+1}-1} (n_{i+1}^{\Gamma_{i+1}-1}-n_{i}^{\Gamma_{i+1}-1})\\
    &= K_{i}n_{i}^{\Gamma_{i}-\Gamma_{i+1}} \frac{\Gamma_{i+1}}{\Gamma_{i+1}-1} (n_{i+1}^{\Gamma_{i+1}-1}-n_{i}^{\Gamma_{i+1}-1})
\end{split}
\end{align}

\noindent where in the last line we have eliminated the yet unknown $K_{i+1}$ by demanding continuity in the pressure $p_i(n_i)=p_{i+1}(n_i)$. The following expression can be solved by a zero search algorithm to find $\Gamma_{i+1}$

\begin{equation}
    \Gamma_{i+1}K_i n_i^{\Gamma_i-\Gamma_{i+1}}=\frac{\mu_{i+1}-\mu_i}{n_{i+1}^{\Gamma_{i+1}}-n_{i}^{\Gamma_{i+1}}}(\Gamma_{i+1}-1).    
\end{equation}

\noindent $K_{i+1}$ and $c_{i+1}$ are then obtained from continuity in pressure and chemical potential, respectively. We will also need the pressure at the start of the next polytrope for the next iteration, which is given by the polytropic equation.

\begin{align}
    \begin{split}
   K_{i+1} &= n_{i}^{-\Gamma_{i+1}}P_i \\ 
   c_{i+1} &= \mu_{i+1}-\frac{K_{i+1}n_{i+1}^{\Gamma_{i+1}-1}          \Gamma_{i+1}}{\Gamma_{i+1}-1} \\ 
   P_{i+1} &= K_{i+1}n_{i+1}^{\Gamma_{i+1}}.
    \end{split}
\end{align}

\noindent For the first polytrope, $K_0$ has to be calculated once from the first data point of $\mu$ and $n$ via

\begin{equation}
    K_0 = (\mu_0-c_0)\frac{\Gamma_0-1}{\Gamma_0}n_0^{1-\Gamma_0}
\end{equation}

\noindent Finally, the speed of sound is given by

\begin{equation}
    c_{s,i}^2 = \frac{dp}{d\varepsilon} = \frac{dp}{dn}\left(\frac{d\varepsilon}{dn}\right)^{-1} = \frac{\Gamma_i p}{p+\varepsilon}.
\end{equation}

\noindent As is visible, this implies usually that the speed of sound is discontinuous, when switching from one polytrope to the next.

In principle, this could again be cured, by combining more data points to fix the derivatives, introducing new parameters. This can be escalated, however this will always yield a discontinuity at latest at the $n$'th derivative for $n$ data points. This could only be avoided by a global fit, using a continuous function. This would, however, also eliminate phase transitions. Eventually, there is no optimal choice with having only a finite number of data points. Since the TOV equations do not depend on the speed of sound, and do not require it to be continuous, but only the (guaranteed) continuity and compatibility of the EoS, polytropes are the minimal solution to guarantee this feature. The slight kinks are integrated over, and thus an overestimation and underestimation by the kinked solutions will average, to some degree, out. Hence, we choose this option here, following \cite{Kurkela:2014vha, Dengler:2021qcq, Pitz:2024xvh}. The stability analysis in \Cref{ss:stability} uses only integrated quantities and initial conditions, and is therefore not sensitive to local kinks in the EoS.

We apply this framework for both the ordinary matter EoS and the dark matter EoS.

\subsection{Additional results}\label{a:res}

\begin{figure}[ht]
\begin{center}
    \includegraphics[width = 0.99\textwidth]{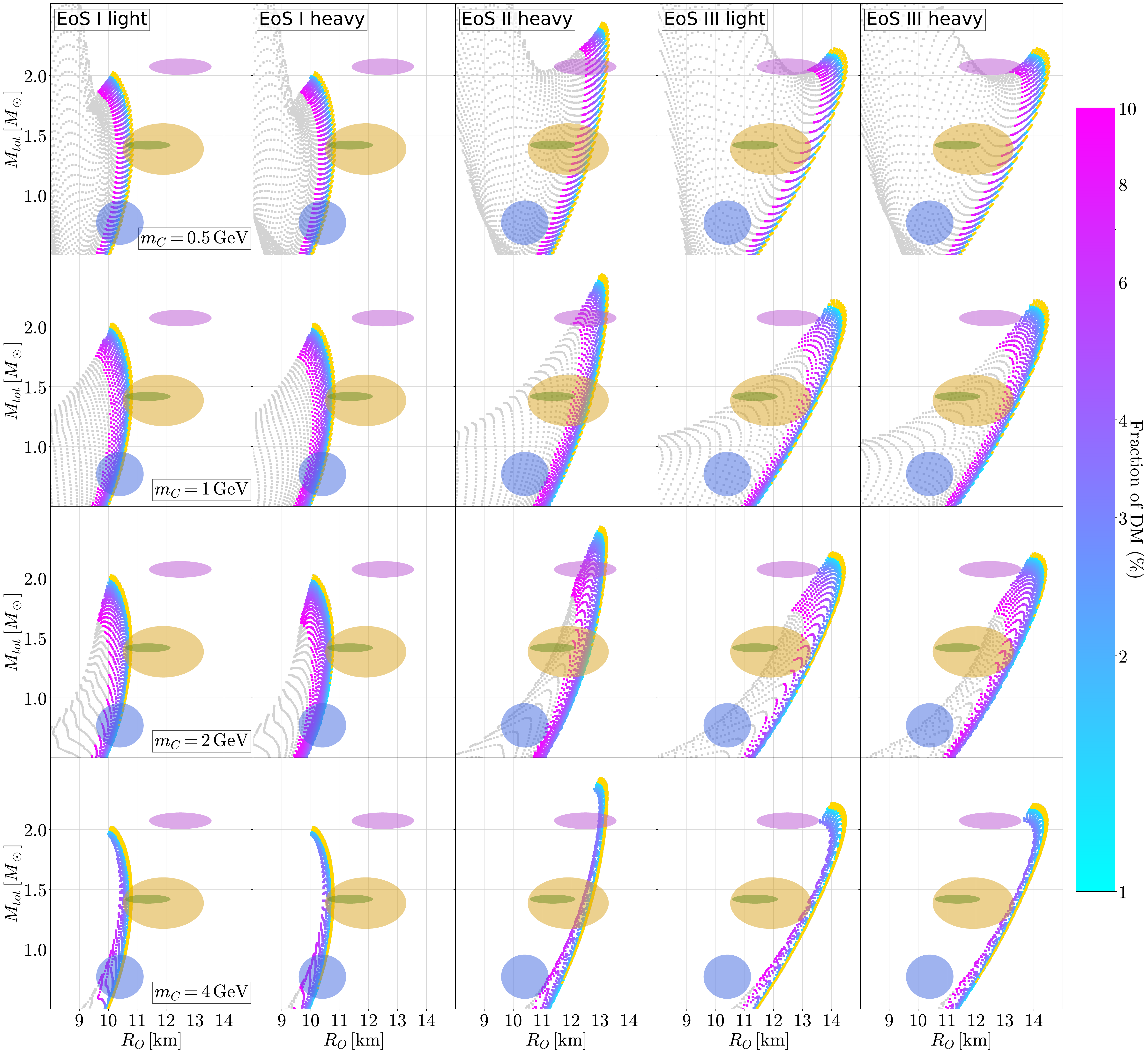}
    \caption{Same as \Cref{plot:mr}, but for the remaining combinations of equations of state.}
    \label{plot:app_m_r}
\end{center}
\end{figure}

\begin{figure}[ht]
\begin{center}
    \includegraphics[width = 0.99\textwidth]{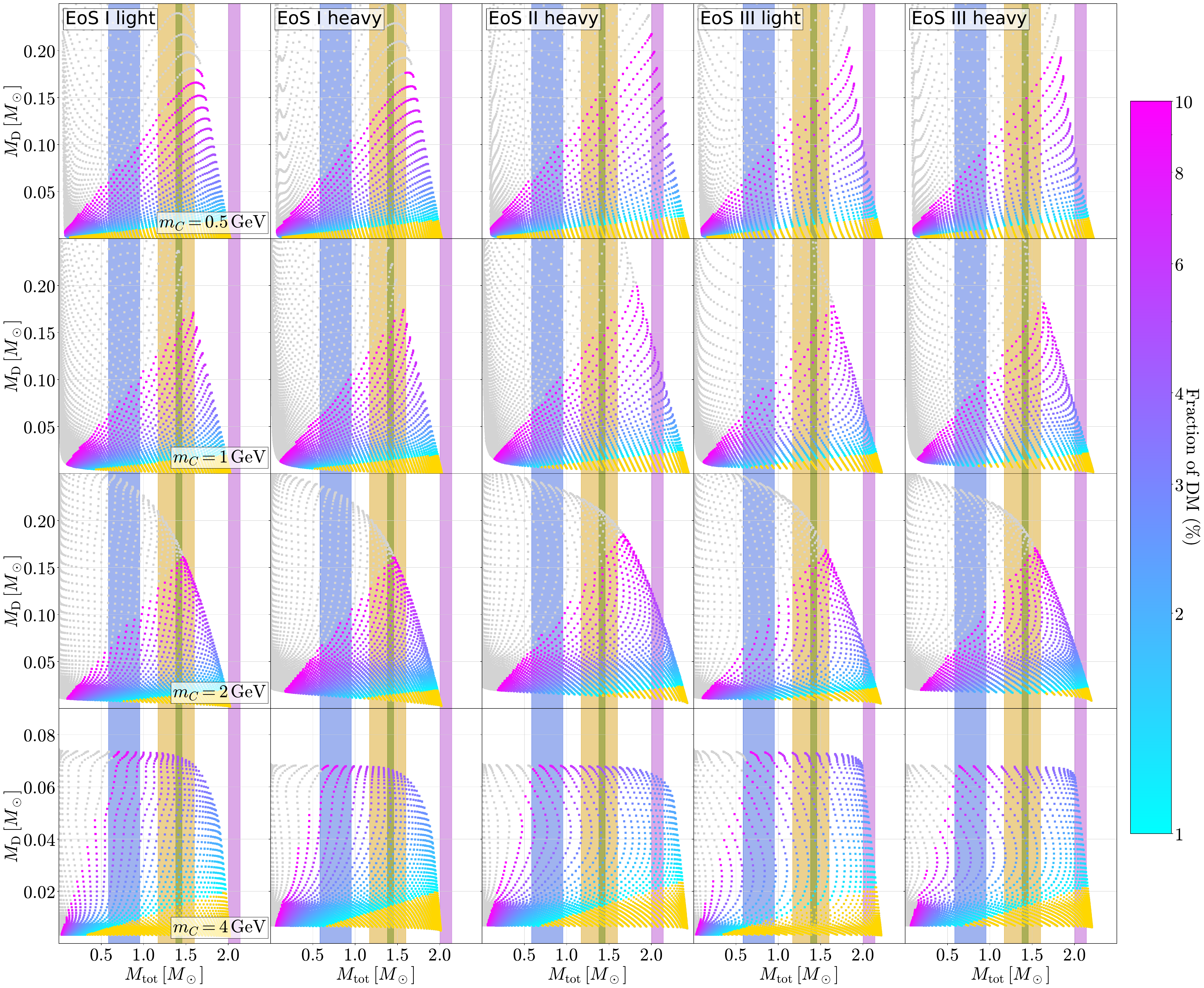}
    \caption{Same as the left of \Cref{plot:mm_and_rr}, but for the remaining combinations of equations of state.}
    \label{plot:app_m_m}
\end{center}
\end{figure}

\begin{figure}[ht]
\begin{center}
    \includegraphics[width = 0.99\textwidth]{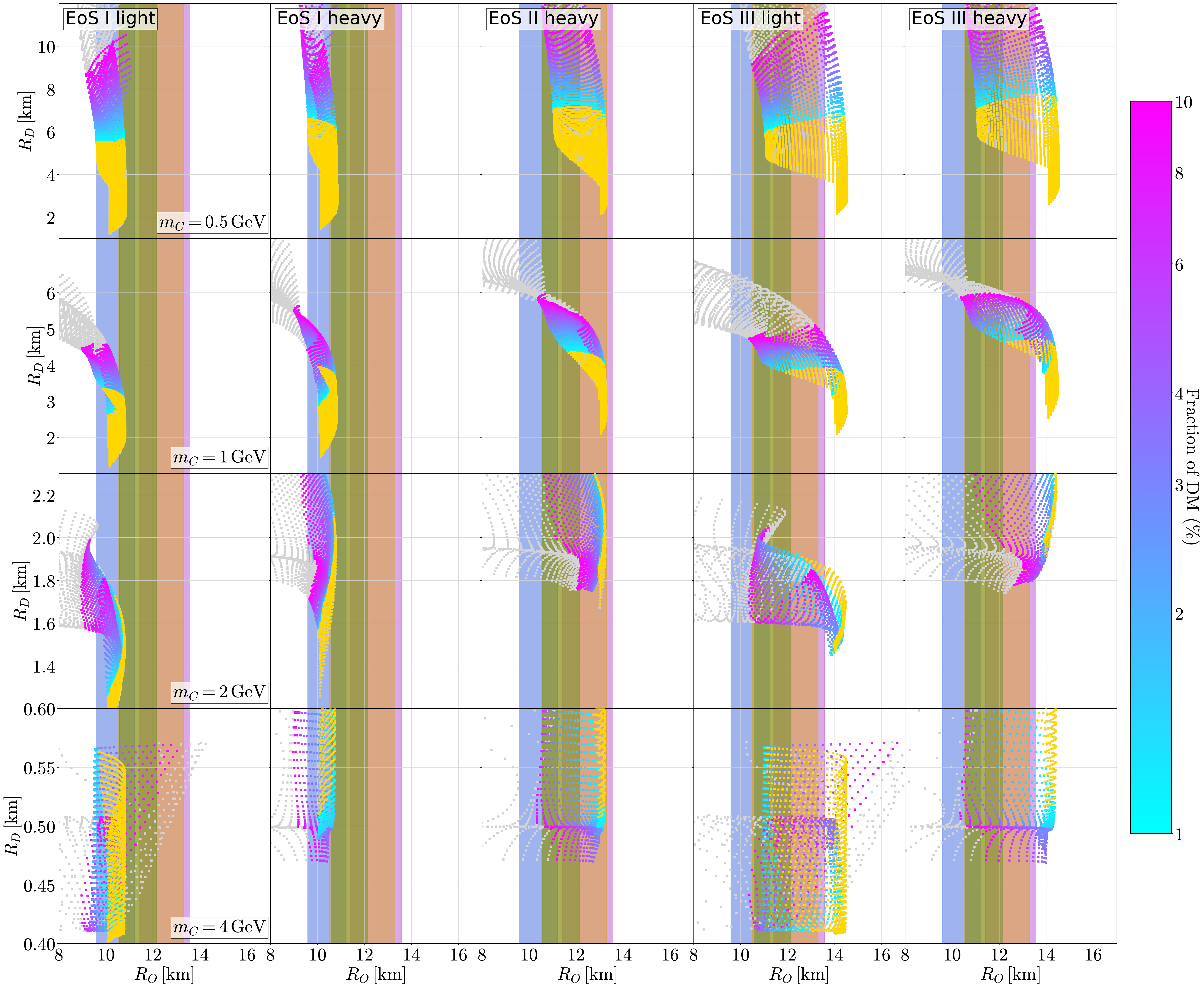}
    \caption{Same as the right of \Cref{plot:mm_and_rr}, but for the remaining combinations of equations of state.}
    \label{plot:app_r_r}
\end{center}
\end{figure}

\begin{figure}[ht]
\begin{center}
    \includegraphics[width = 0.99\textwidth]{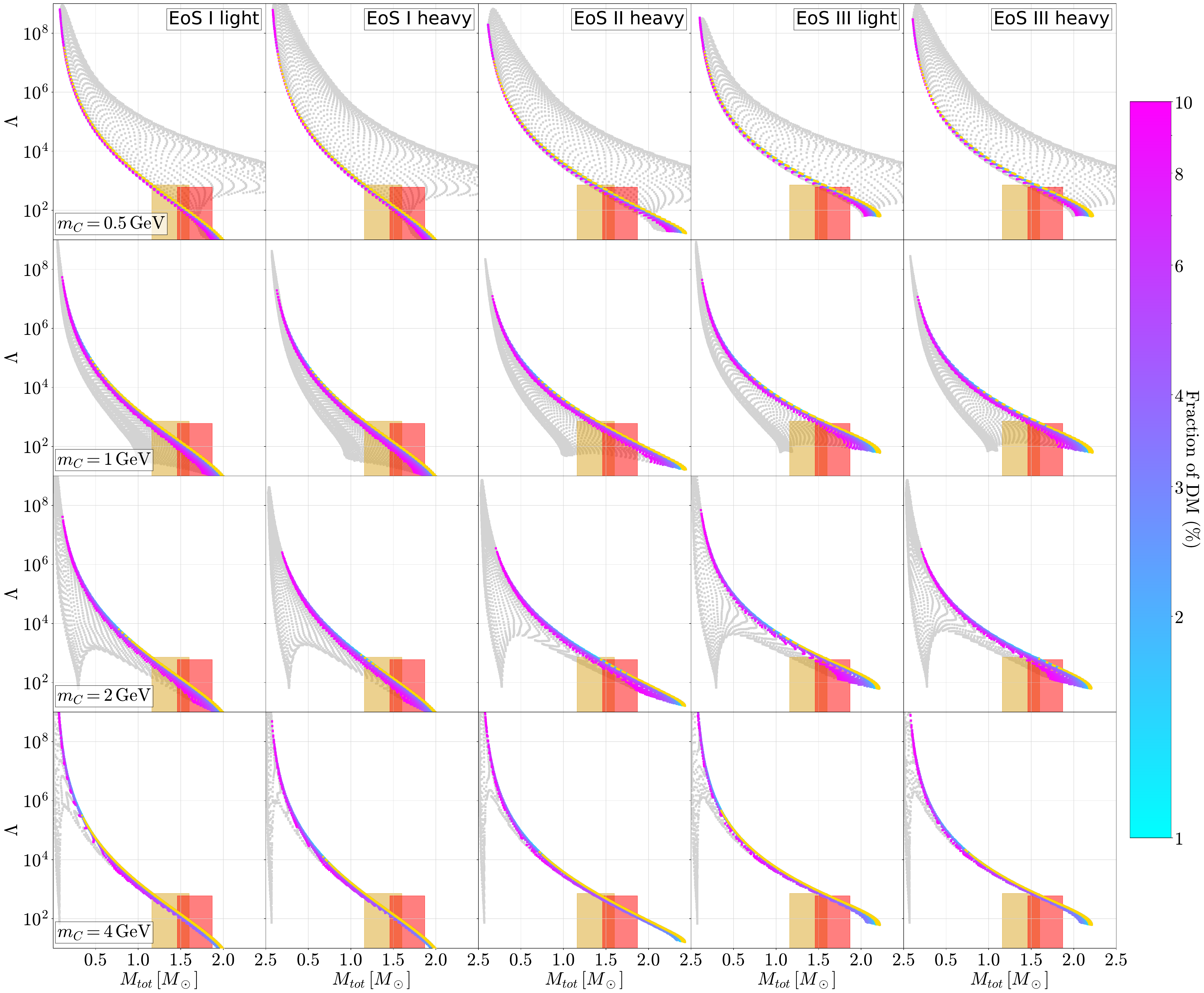}
    \caption{Same as \Cref{plot:lambda}, but for the remaining combinations of equations of state.}
    \label{plot:app_lambda_m}
\end{center}
\end{figure}

In this appendix, we show additional results for the masses, radii, and tidal deformabilities. \Cref{plot:app_m_r} shows the mass-radius relations for the remaining combinations of EoS. First of all, we see that the shape of the mass-radius relation depends mostly on the ordinary matter EoS employed. This is expected as we restrict ourselves to <10\% dark matter here. As a result, whether or not the mixed stars are compatible with the measurements also mostly depends on the ordinary matter EoS. We see, however, that the addition of dark matter can significantly change both the mass and the radius. We further see that the impact of the light dark matter and heavy dark matter EoS are very similar. We see a similar picture in the composition of the individual masses in in \Cref{plot:app_m_m} but find that the light EoS can result in slightly higher dark matter masses. \Cref{plot:app_r_r} shows the dark matter radius versus the ordinary matter radius. At low dark matter candidate masses, the impact of the light and heavy EoS are very similar. For large candidate masses, however, where the dark matter forms a compact core, the results for the light and heavy ensemble differ. The dark matter EoS results in much larger dark matter radii. Finally, in \Cref{plot:app_lambda_m}, we display the tidal deformability for the remaining combinations of EoS, which does not seem to be dependent on the dark matter EoS employed and mostly depends on the mass of the dark matter candidate.

\begin{figure}[ht]
\begin{center}
    \includegraphics[width = 0.99\textwidth]{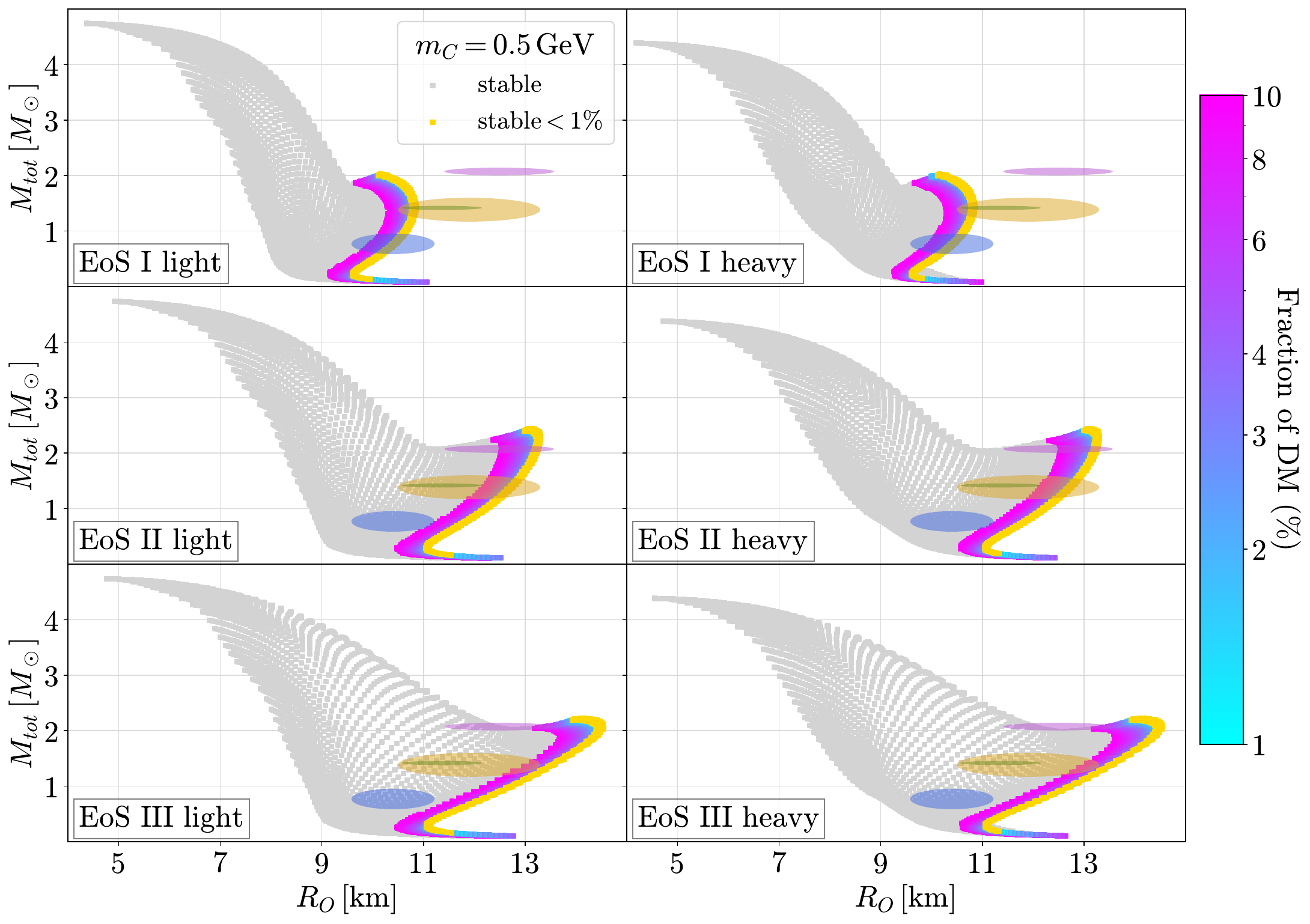}
    \caption{Same as \Cref{plot:mr}, but for all combinations of equations of state using $m_C=0.5\,$GeV using a larger range to show the non-standard compact object solutions.}
    \label{plot:app_non_standard}
\end{center}
\end{figure}

In \Cref{plot:app_non_standard} we show the mass-radius relation with $m_C=0.5\,$GeV in a larger range to investigate solutions with large amounts of dark matter. The dark matter component in this case forms a large and massive dark matter halo which can reach a mass $>4\,M_\odot$. These solutions do not resemble neutron stars at all, but could be interpreted as heavy non-standard compact stellar objects with very large masses at ordinary matter radii similar to neutron stars and smaller.


\bibliographystyle{JHEP}
\bibliography{bibliography.bib}

\end{document}